\documentstyle[12pt,preprint,aps,epsf]{revtex}
\def\thalf{{\textstyle{\frac{1}{2}}}}

\begin{document}
\preprint{\normalsize NUC-MINN-02/2-T}
\title{Dileptons and Photons from Coarse-Grained Microscopic Dynamics and 
       Hydrodynamics Compared to Experimental Data }
\author{{\bf P. Huovinen$^1$, M. Belkacem$^2$, P. J. Ellis$^{1}$ 
         and J. I. Kapusta$^{1}$} \vspace*{0.1in} \\ 
{\it $^{1}$School of Physics and Astronomy, University of Minnesota}\\
{\it Minneapolis, MN 55455, USA} \\ 
{\it $^{2}$Laboratoire de Physique Quantique, Universit\'e Paul Sabatier}\\
{\it 31062 Toulouse Cedex, France} }
\date{\today}

\maketitle

\thispagestyle{empty}

\begin{abstract}
Radiation of dileptons and photons from high energy nuclear collisions
provides information on the space-time evolution of the hot dense
matter produced therein. We compute this radiation using relativistic
hydrodynamics and a coarse-grained version of the microscopic event
generator UrQMD, both of which provide a good description of the
hadron spectra.  The currently most accurate dilepton and photon
emission rates from perturbative QCD and from experimentally-based
hadronic calculations are used.  Comparisons are made to data on
central Pb-Pb and Pb-Au collisions taken at the CERN SPS at a beam
energy of 158 A GeV. Both hydrodynamics and UrQMD provide very good
descriptions of the photon transverse momentum spectrum measured
between 1 and 4 GeV, but slightly underestimate the low mass spectrum
of $e^+e^-$ pairs, even with greatly broadened $\rho$ and $\omega$
vector mesons. Predictions are given for the transverse momentum
distribution of dileptons.
\end{abstract}

\noindent PACS {25.75.-q, 13.85.Qk}

\section{Introduction}

The primary rationale for colliding large nuclei at high energy is to
produce high-density matter in as large a volume as possible for as
long as possible.  The energy density of atomic nuclei sets the scale
at 0.15 GeV/fm$^3$. The energy density in neutron stars may reach
two to ten times that number.  The energy density for the phase transition
or crossover from hadronic to quark and gluon degrees of freedom is
about ten times that number.  In terms of temperature, the
phase transition or crossover should occur between 150 and 200 MeV in
net-baryon-free matter.  A first principles theoretical description of
the space-time evolution of the produced matter is not yet possible.
It is very much more complicated than lattice quantum chromodynamics
(QCD) calculations of the thermodynamic properties. There are only a
few well-developed dynamical models of high energy nuclear collisions
which provide a space-time description of matter.  Chief among these
are UrQMD (Ultra relativistic Quantum Molecular Dynamics)
\cite{urqmd1,urqmd2} and relativistic, non-viscous, perfect fluid
dynamics, simply referred to as hydrodynamics 
\cite{soll}.  UrQMD is restricted to hadronic degrees of freedom, while
hydrodynamics can, as here, employ an equation of state which includes
both quark-gluon and hadronic phases. Both of these models require
input that is not obtainable from experimental data on hadron-hadron
collisions and is not yet computable from QCD.  In the case of UrQMD
one must estimate unmeasured or unmeasurable cross sections, usually
involving hadron resonances like the $\Delta$ or the $\rho$.  There
are also difficulties in dealing with hadron-hadron collisions at high
density, for instance in the treatment of off-mass-shell effects and
simultaneous collisions of more than two particles. In the case of
hydrodynamics one does not know the initial conditions very well, and
there is some difficulty in determining when the transition from local
thermal equilibrium to free-streaming particles occurs.  Nevertheless,
sufficient work has been done on these problems to achieve good,
theoretically acceptable descriptions of the abundances of different
types of hadrons and their momentum distributions for heavy ion
collisions at the CERN SPS using both approaches
\cite{urqmd1,soll,Sollfrank99}.

The distribution of observed hadrons does not readily provide
information on the state of matter prior to their last scattering.  It
was suggested long ago that the emission of direct photons (those not
arising from the decay products of unstable hadron resonances after
freezeout) and dileptons ($e^+e^-$ and/or $\mu^+\mu^-$) would provide
information on the temperature and baryon density of the evolving
matter.  The reason is that these electromagnetic probes generally
have mean free paths that are much greater than the spatial dimensions
of the system.  Calculations of the emission rates have become very
sophisticated, and there is now essentially a consensus on their
magnitudes.  For recent reviews see Peitzmann and Thoma \cite{pt} and
Rapp and Wambach \cite{rwrev}.  However, in order to use these thermal
emission rates one must know the local temperature and baryon chemical
potential as functions of space and time.  This is straightforward to
obtain in hydrodynamics.  To obtain this information in a microscopic
event generator, such as UrQMD, one must accumulate an ensemble of
collision events and then determine the local variables via
coarse-graining.  We have done this and will report the results in
this paper.  The system we focus on is central Pb-Pb collisions at a
laboratory beam energy of 158 A GeV, equivalent to a center-of-mass
energy of 17 A GeV.  The reason we focus on this system is that there
is a relative wealth of data taken at the CERN SPS for hadrons,
photons, and dileptons.  This is not the case at the RHIC
(Relativistic Heavy Ion Collider) at Brookhaven National Laboratory
where electromagnetic spectra have not yet been measured and only
hadron spectra are available.

In Sec. II we compare the space-time evolution of the nuclear
collision obtained from hydrodynamics to that obtained from a
coarse-graining of UrQMD.  In Sec. III we compare the distributions of
negatively charged hadrons and net protons from the coarse-grained
UrQMD to experimental data, just to be sure that the coarse-graining
has not severely distorted the model predictions.  In Secs. IV and V
we compare the results of these two dynamical models to experimental
data on photons and dileptons, respectively.  We summarize our results
in Sec. VI.

\section{Space-Time Evolution}

The initial conditions for the hydrodynamic calculation cannot be
framed in terms of two colliding nuclei since the primary particle
production is a non-adiabatic process which is not describable by
perfect fluid dynamics.  Instead the hydrodynamic calculation begins
with estimated initial conditions soon after impact which are
constrained to reproduce the final observed hadron spectra.  The UrQMD
does begin with the two nuclei approaching each other; strings are
formed and decay, producing particles that subsequently rescatter
according to hadron-hadron cross sections determined independently. We
do not attempt to coarse-grain the initial stage of the collision when
many strings are present.  Rather, we begin after most or all of the
strings have decayed.

The coarse-graining of 100 UrQMD central Pb-Pb collisions is performed
in terms of the following variables: proper time $\tau =
\sqrt{t^2-z^2}$, space-time rapidity $\eta = \thalf
\ln\left[(t+z)/(t-z)\right]$, cylindrical radius $r$, and azimuthal
angle $\phi$.  The origin of the time axis is chosen to be the time
when the colliding nuclei overlap. A grid of small space-time cells is
set up. Using an ensemble of 100 events, the average total momentum,
energy, and net baryon number in each cell is easily found. This
allows us to determine the velocity of the center-of-momentum, the
energy density, and the baryon density for each cell. A table with the
equation of state of an ideal hadron gas containing the same hadronic
degrees of freedom as UrQMD is then used to determine the equivalent
temperature and baryon chemical potential for each cell.  This
equation of state is very similar to that obtained in UrQMD without
strings \cite{mohamed}.  (Inclusion of strings violates the principle
of detailed balance unless the back reactions, involving the collision
of many particles to reform a string, are included and this is not
technically possible at present.) This procedure is justified in our
case since the coarse-graining is done after most or all strings have
already disappeared, as already mentioned. As shown in the next
section this procedure does not severely distort the final hadron
spectra.

Our hydrodynamical model has been described in detail in Refs.
\cite{soll,Huovinen99}. To keep our dilepton results comparable
with the results shown in~\cite{pasprak}, we use the same equation of
state with a first order phase transition to a quark-gluon plasma at a
temperature of $T_c = 200$ MeV, in net baryon free matter, which is at
the high end of the range of expected critical temperatures.  The
parametrization of the initial state is also the same, namely, IS\,1
from Ref. \cite{Huovinen99}.

The version of hydrodynamics applied here does not assume boost
invariance.  Therefore a comparison of the space-time evolution with
UrQMD has the ambiguity of fixing the zero of time.  In the following
figures and discussion we have taken $t=0$ to be the time at which we
initialize hydrodynamics and estimate that this corresponds to 0.5
fm/c after the complete overlap of the two nuclei in UrQMD.

\begin{figure}[t]
 \setlength{\epsfxsize}{5.5in}
  \centerline{\epsffile{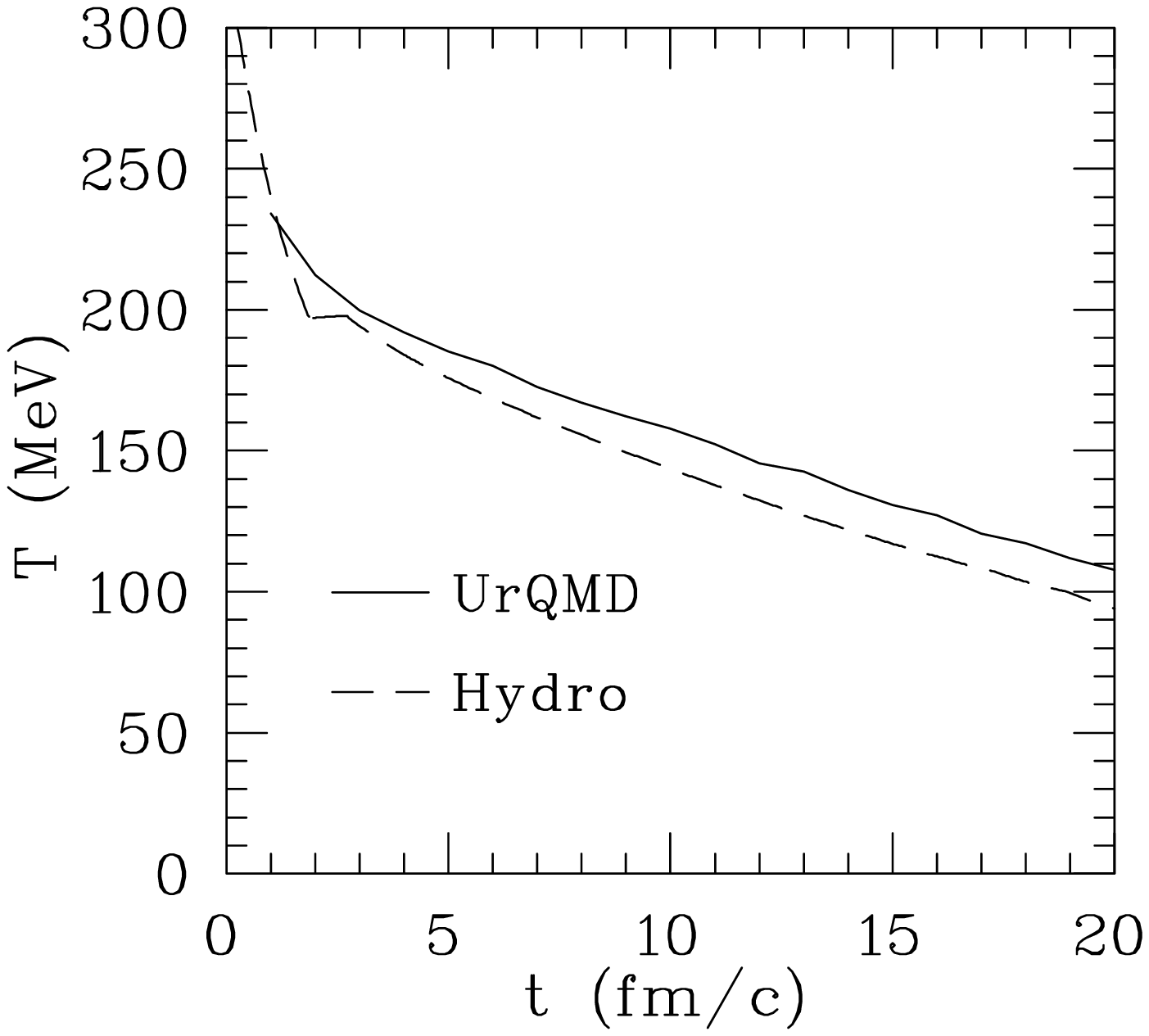}}
\vskip1cm
\caption{The temperature as a function of local time at 
the origin in the UrQMD model (solid curve) and the hydrodynamic model 
(dashed curve).}
 \label{fig1}
\end{figure}

\begin{figure}[t]
 \setlength{\epsfxsize}{5.5in}
  \centerline{\epsffile{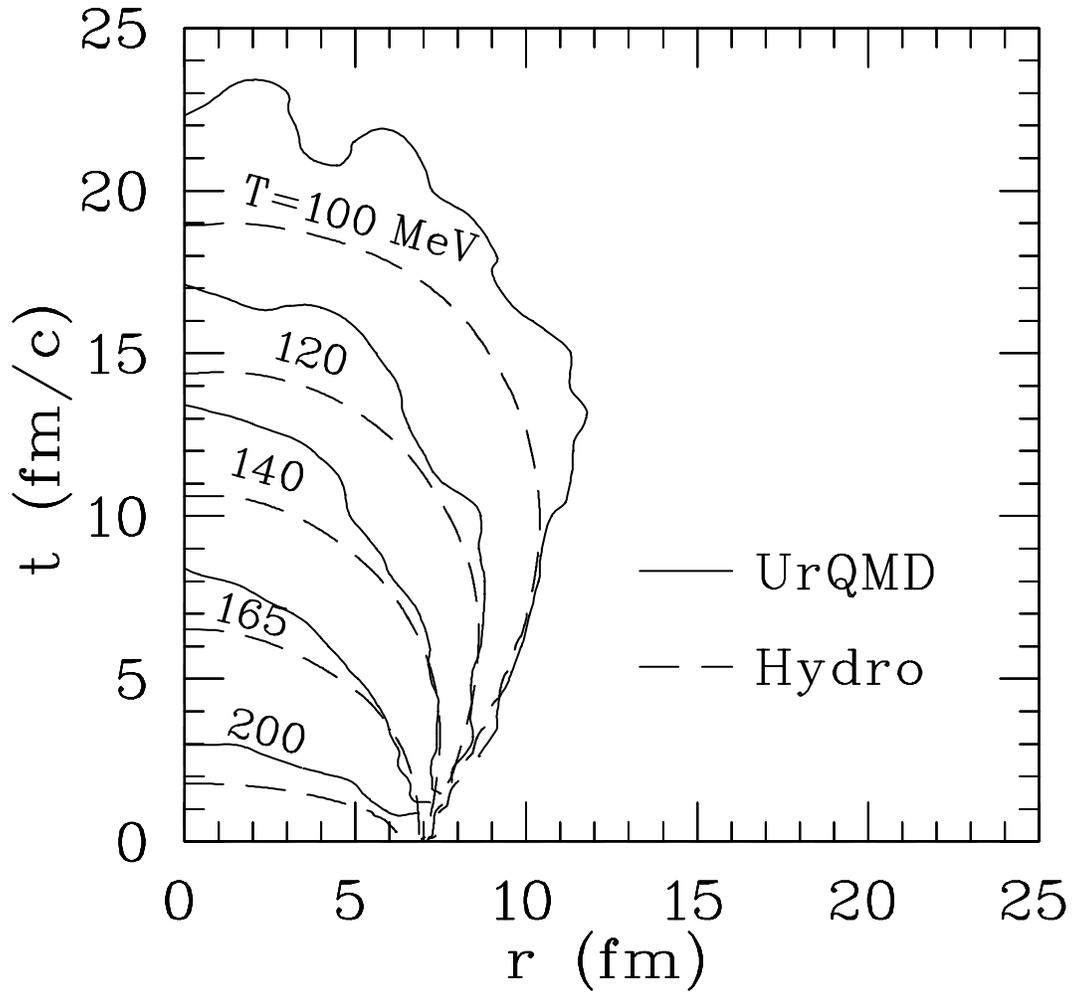}}
\vskip1cm
 \caption{Temperature contours in local time
          and cylindrical radius in the central transverse plane for the UrQMD
          model (solid curves) and the hydrodynamic model (dashed curves).}
 \label{fig2}
\end{figure}

When comparing hydrodynamics to the coarse-grained UrQMD we shall
focus for the most part on the central ($z=0$) transverse plane.  In
Fig. \ref{fig1} we plot the temperature as a function of local time in
a small volume centered at the origin. The effect of the phase
transition on the equation of state used by hydrodynamics can be seen
at a time around 2 fm/c. The equation of state from UrQMD uses
hadronic degrees of freedom only; thus high temperature here means a
superheated resonance gas.  The two curves in Fig. \ref{fig1} are
remarkably similar, although the evolution in UrQMD is somewhat slower
in the sense that the matter takes a longer time to cool.  Cooling in
the transverse plane is displayed as a contour plot in
Fig. \ref{fig2}.  It shows the contours of constant temperature in a
slice of the $t-r$ plane centered at $z=0$.  Note that the small
wiggles arise from the coarse graining of the UrQMD with a finite
number of events. The two models yield curves which are qualitatively
similar, but once again it is apparent that the matter described by
UrQMD takes longer to cool.  The most likely explanation is that the
matter evolution described by UrQMD dynamics includes the effects of
viscosity and heat conduction which will slow its expansion and
cooling.

\begin{figure}[t]
 \setlength{\epsfxsize}{5.5in}
  \centerline{\epsffile{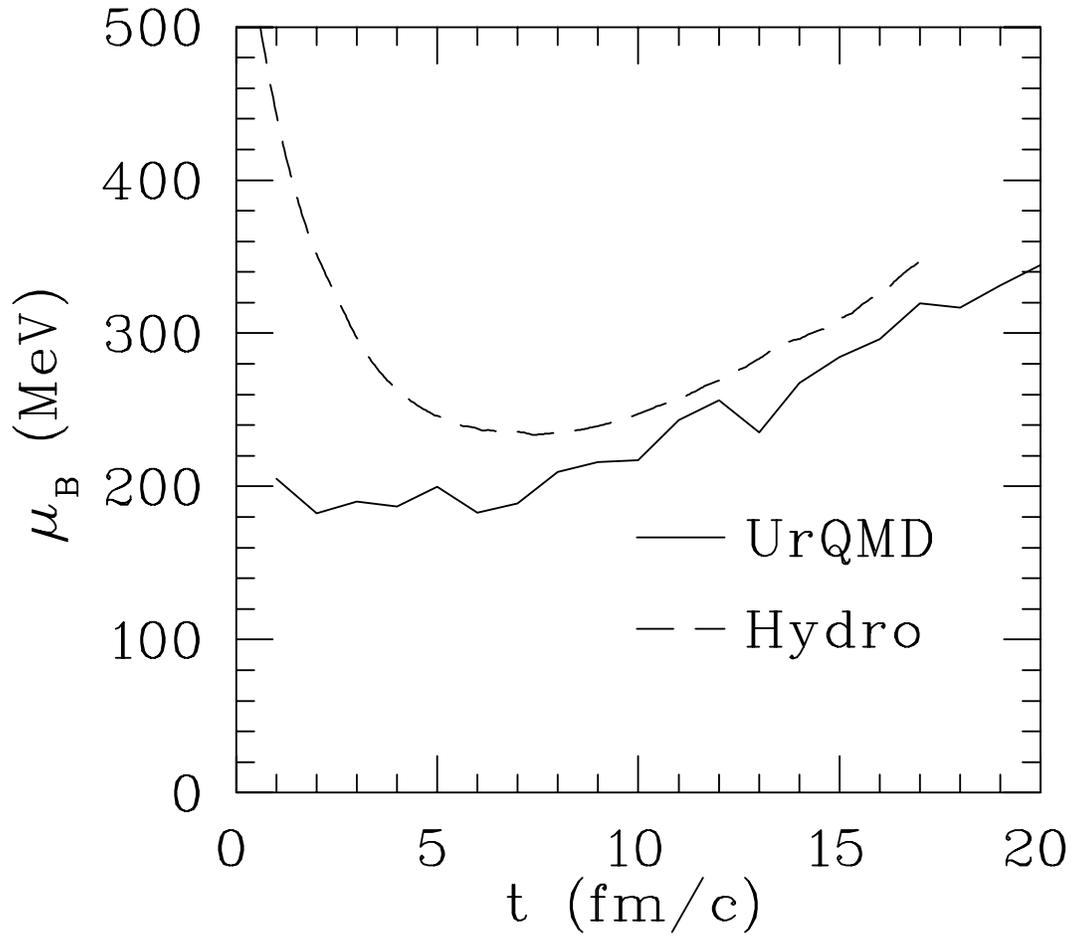}}
\vskip1cm
\caption{The time evolution of the baryon chemical potential at the origin 
in the UrQMD model (solid curve) and the hydrodynamic model (dashed 
curve).}
 \label{fig3}
\end{figure}

The baryon chemical potential evolution in a small volume centered at
the origin is plotted in Fig. \ref{fig3}.  Due to the finite number of
events in UrQMD the maximum value of the chemical potential does not
occur precisely at the origin. Thus the discrepancy between UrQMD and
hydrodynamics in the first few fm/c is more apparent than real. After
5 fm/c the evolution is evidently quite similar in the two models.

\begin{figure}[t]
 \setlength{\epsfxsize}{5.5in}
  \centerline{\epsffile{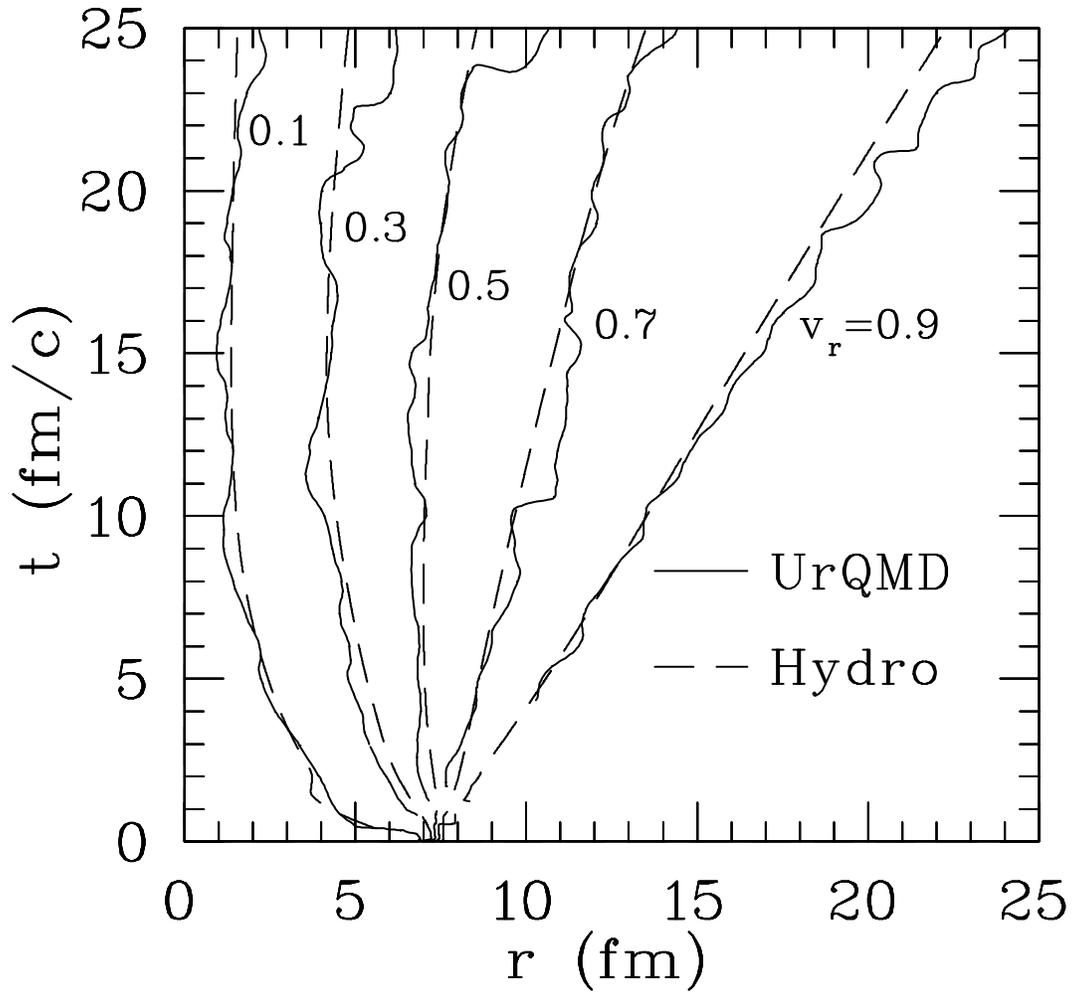}}
\vskip1cm
\caption{Contours of equal radial flow velocity as a function of local time
         and cylindrical radius in the central transverse plane for the UrQMD
         model (solid curves) and the hydrodynamic model (dashed curves).}
\label{fig4}
\end{figure}

\begin{figure}[t]
 \setlength{\epsfxsize}{5.5in}
   \centerline{\epsffile{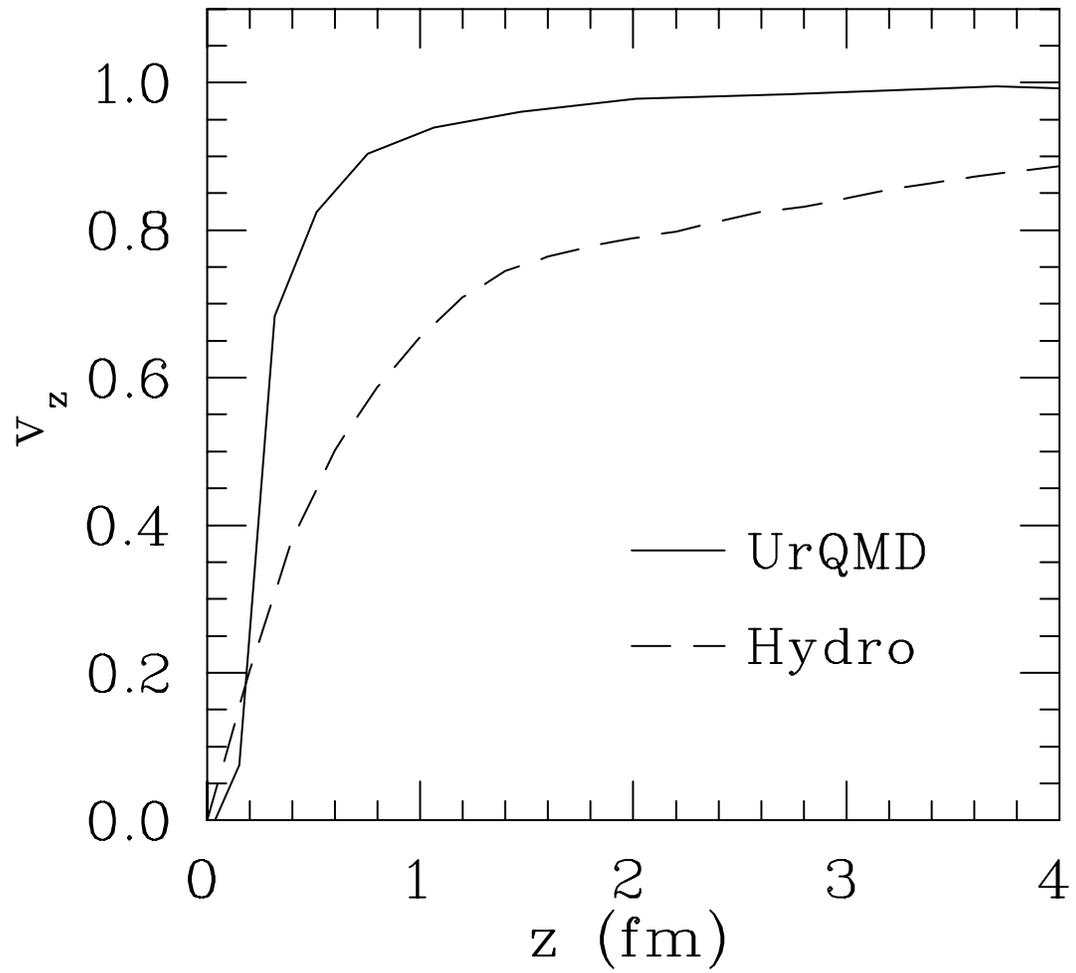}}
\vskip1cm
\caption{The initial velocity distribution along the beam axis
in the UrQMD model (solid curve) and the hydrodynamic model 
(dashed curve).}
 \label{fig5}
\end{figure}

Contours of the radial flow velocity in the central plane are
displayed in Fig. \ref{fig4}.  Apart from fluctuations in the UrQMD
results arising from using a finite number of events in the
coarse-graining, the contours are almost identical. A more striking
difference is seen in the initial\footnote{The reader is advised to
keep in mind the ambiguity in fixing the $t$-axis as explained before.
The UrQMD velocity profile is for $\tau=0.5$ fm/c, whereas the
hydrodynamic profile is for $\sqrt{(t+0.5)^2-z^2}=0.5$ fm/c, where
time $t$ is measured from the beginning of the hydrodynamical
evolution.}  velocity distribution along the beam (or $z$) axis,
displayed in Fig.~\ref{fig5}. Whereas in UrQMD this velocity
distribution is essentially an outcome of the string dynamics, in
hydrodynamics it is an assumed input.  The UrQMD velocity profile
closely resembles a boost-invariant velocity profile
$v=z/\sqrt{\tau^2+z^2}$ with time $\tau = 0.35$ fm/c except in the
region close to the origin, $|z| < 0.4$ fm, where the flow velocity is
less than in the boost invariant case. Conversely the hydrodynamic
profile is close to the boost invariant one for $\tau=1.0$ fm/c in the
region $|z|<1.0$ fm, whereas at large $z$ the velocity is smaller than
that of a boost-invariant profile. Thus the deviations from boost
invariant flow are qualitatively different and the differences in
velocity profiles cannot be due to the ambiguity in defining the
initial time. It should also be remembered that even if the
longitudinal velocity profiles in both models resemble boost invariant
flow, the matter distributions do not since they peak at $z=0$.

We have seen that the two models show a remarkable similarity in the
temperature, baryon chemical potential, and transverse flow results
near the origin and in the central transverse plane.  This is most
likely due to the (approximate) boost invariance, an essential feature
in high energy collisions.  In UrQMD it is a consequence, in some
sense, of string dynamics.  In hydrodynamics it is a combination of
relativity and the assumed initial conditions.  The small differences
are most likely attributable to the fact that the microscopic model
includes in an essential way the physical effects of shear and bulk
viscosity and heat conduction.  The similarities are not so striking
in the longitudinal direction, which is probably a consequence of the
way that finite energy effects are actually implemented in the two
dynamical models (exact boost invariance occurs only at infinite beam
energy).  Nevertheless the similarities are great enough that one may
conclude that the UrQMD results in something akin to local thermal
equilibrium, at least insofar as we have probed it with the
coarse-graining procedure.  It would be very interesting and
worthwhile to carry out fluid dynamic calculations with the initial
conditions obtained from UrQMD, both with and without viscosity and
heat conduction included.  Such studies are the subject of another
paper.

\section{Hadron Spectra}

The initial conditions of hydrodynamics are chosen so that experimental
hadronic spectra for Pb-Pb collisions at 158 A GeV are relatively well 
reproduced~\cite{Sollfrank99,Huovinen99}.
The results of UrQMD also reproduce the experimental hadronic spectra
quite well, apart from hyperons containing two or three strange
quarks which may signal interesting physics not contained in the
model \cite{urqmd1,urhad,urstr}.

To investigate whether the coarse-graining distorts the hadronic spectra
we have calculated the hadronic spectra from the coarse-grained UrQMD
temperature, chemical potential, and flow velocity fields in the same way
as in hydrodynamics\footnote{The hadronic equation of state used in 
   the hydrodynamic calculations is essentially identical to the 
   equation of state obtained from running UrQMD without strings in
   a large box.}.
When coarse-graining is done on a microscopic model like UrQMD one faces 
exactly the same issue as in hydrodynamics: when to stop and compute 
the hadronic spectra. We have generally taken a range of freeze-out 
temperatures, namely 140, 120, and 100 MeV, and results will be displayed
for these cases. For comparison the default temperature at which the
hydrodynamic evolution ceases and free-streaming begins was chosen to be
140 MeV, a value determined in earlier studies \cite{Huovinen99}.

\begin{figure}[t]
 \setlength{\epsfxsize}{5.5in}
  \centerline{\epsffile{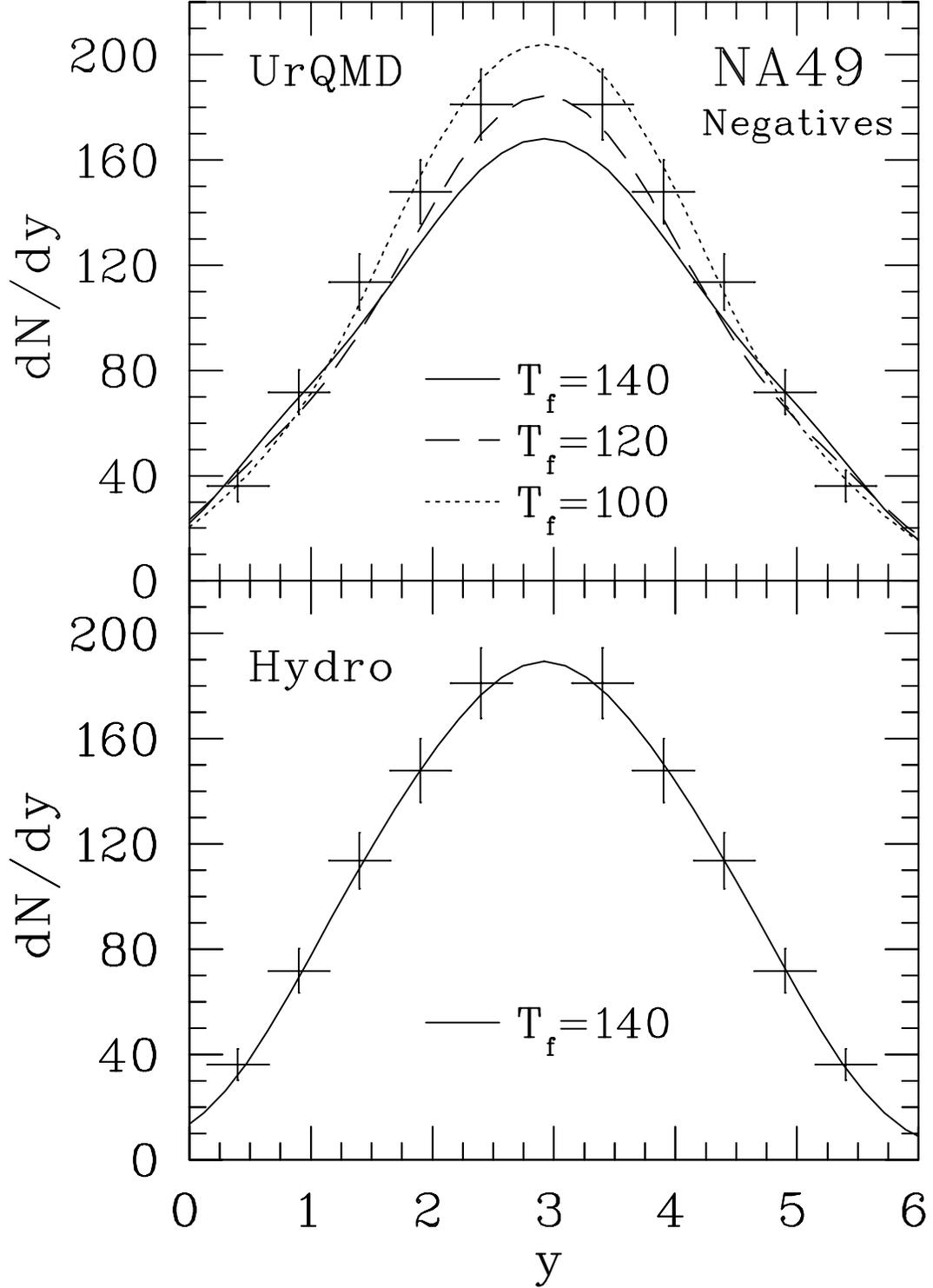}}
\vskip1cm
\caption{The rapidity distribution of negatively charged hadrons. Upper 
         panel: UrQMD for various freeze-out temperatures; lower
         panel: hydrodynamic model; data: NA49 collaboration
         \protect\cite{Jones}.}
  \label{fig6}
\end{figure}

\begin{figure}[t]
 \setlength{\epsfxsize}{5.5in}
  \centerline{\epsffile{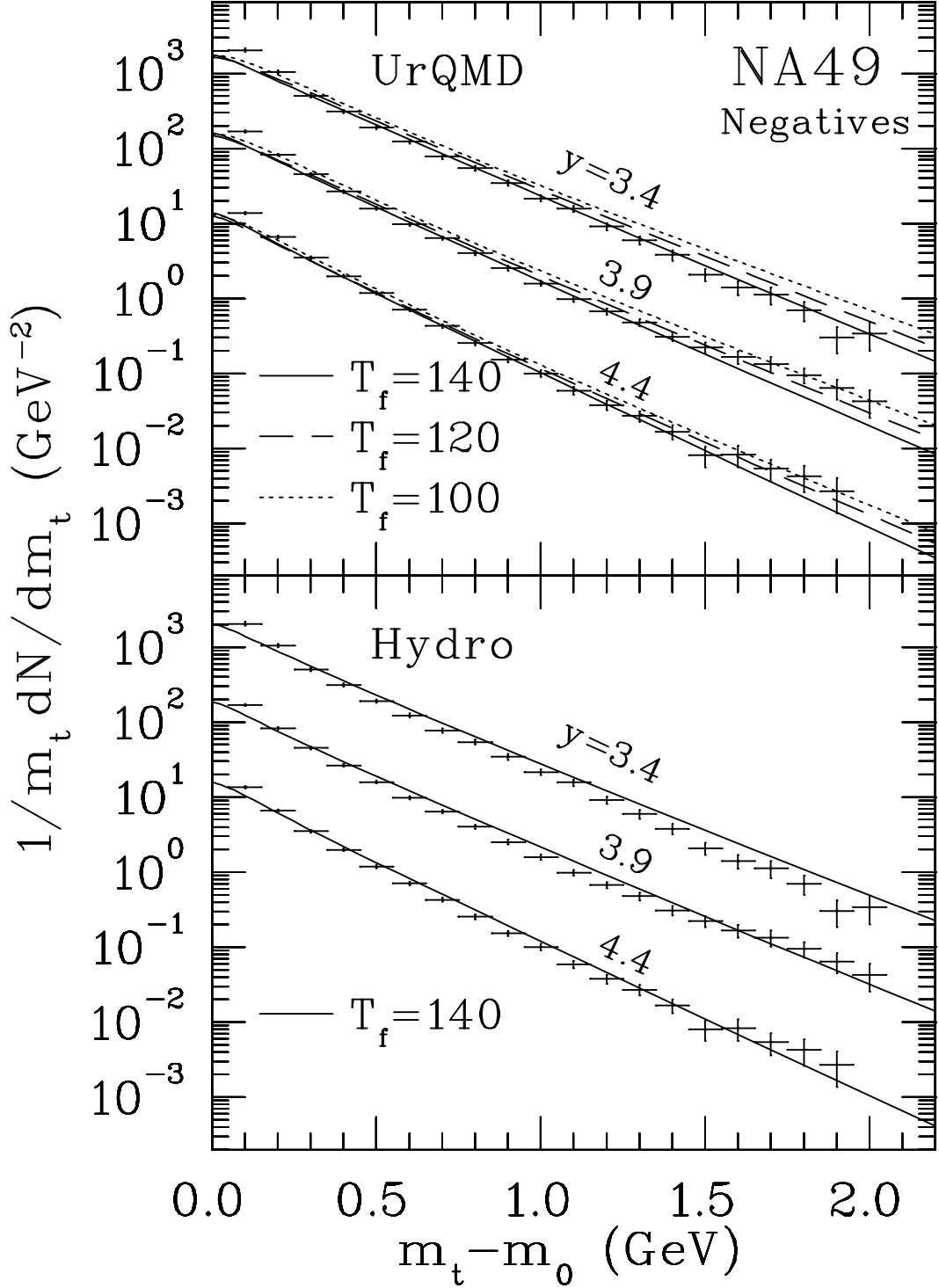}}
\vskip1cm
\caption{The transverse mass distributions of negatively charged hadrons at 
         rapidities of 3.4, 3.9 and 4.4.  Upper panel: UrQMD for
         various freeze-out temperatures; lower panel: hydrodynamic
         model; data: NA49 collaboration~\protect\cite{Jones}.}
\label{fig7}
\end{figure}

In Fig. \ref{fig6} we show the rapidity distribution of negatively
charged hadrons compared to data taken by the NA49 collaboration~\cite{Jones}.
In this and the subsequent figures the top panel shows the UrQMD results
which can be compared with the hydrodynamic result in the bottom
panel. A freeze-out temperature between 120 and 100 MeV in UrQMD
yields the best comparison to the data. The hydrodynamic description
fits the data quite nicely since the initial conditions were tuned to
this end. The transverse mass distributions at laboratory rapidities
of 3.4, 3.9 and 4.4 are shown in Fig. \ref{fig7}.  The UrQMD results
are acceptable for essentially any freezeout temperature in the range
displayed; the hydrodynamic results in the lower panel also agree
with the data.

\begin{figure}[t]
 \setlength{\epsfxsize}{5.5in}
  \centerline{\epsffile{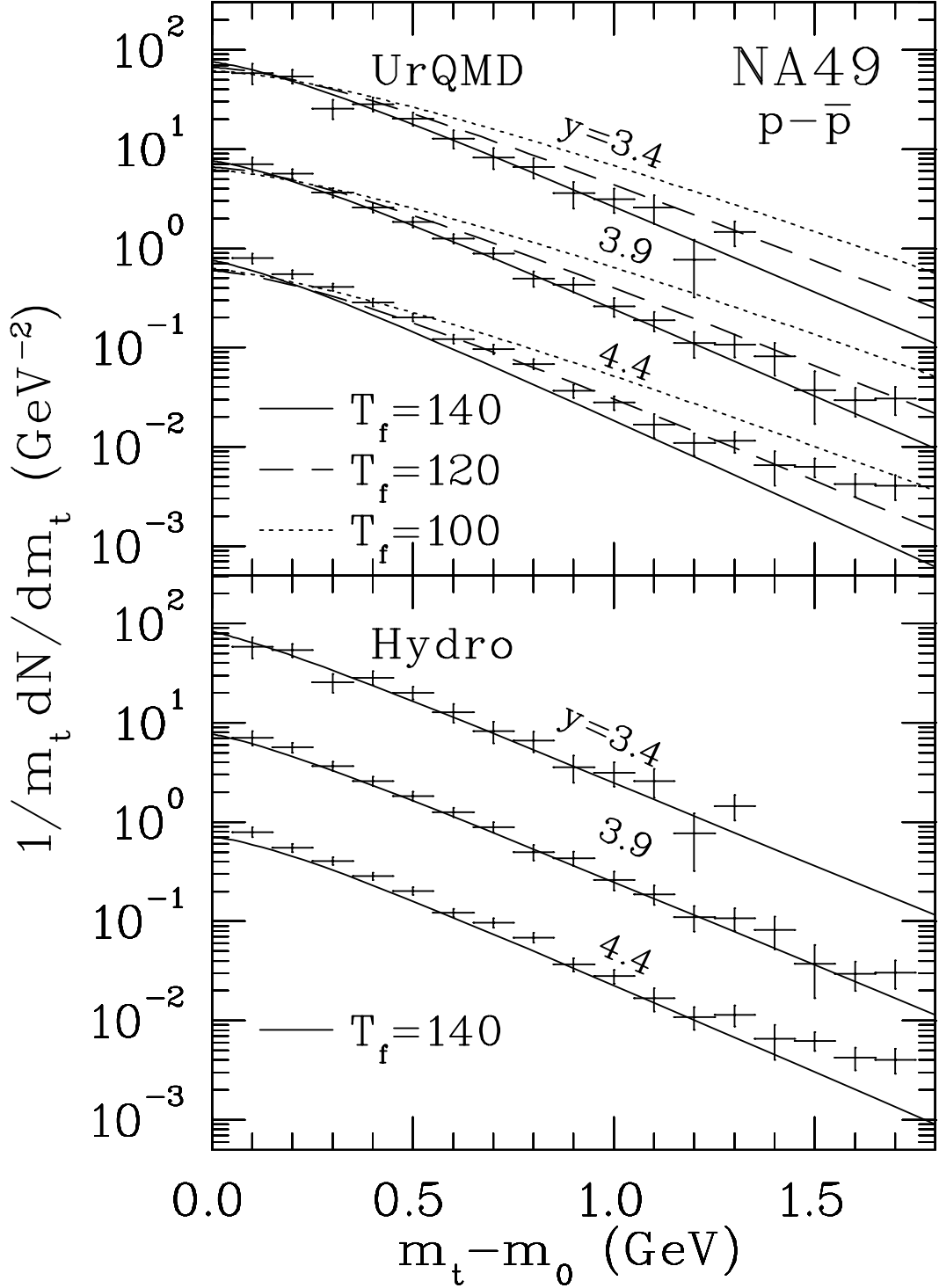}}
\vskip1cm
 \caption{The net proton spectra at rapidities of 3.4, 3.9 and 4.4.  
          Upper panel: UrQMD for various freeze-out 
         temperatures; lower panel: hydrodynamic model;
         data: NA49 collaboration~\protect\cite{Jones}.}
  \label{fig8}
\end{figure}

The net proton (proton minus anti-proton) spectra are displayed in
Fig. \ref{fig8}.  The sensitivity of the UrQMD results to the
freeze-out temperature is greater here than it is for the negative
hadron spectra.  This is understandable since in chemical equilibrium,
with a nonzero baryon chemical potential, the ratio of anti-protons to
protons is exponentially sensitive to the temperature, behaving
approximately as $\exp(-2\mu_B/T)$.  This, together with the increased
transverse flow velocity as the system evolves to lower temperatures,
qualitatively explains the results of Fig. \ref{fig8}.  Although a
freeze-out temperature of around 140 to 120 MeV might seem to be
favored, these results cannot be used to pin down the freeze-out
conditions.  The reason is that the anti-protons are very sensitive to
annihilation processes and to non-equilibrium dynamics in the
relatively dilute stage at the end of the collision.  Still,
considering the extreme assumption of local thermal and chemical
equilibrium used to obtain these results, the outcome is
reasonable. The hydrodynamic results (lower panel) are quite similar
to the UrQMD results when the same freeze-out temperature, 140 MeV, is
employed.

Overall we may conclude that the coarse-graining has not severely
distorted the original UrQMD results which represent the experimental
data well.

\section{Photon Spectra}

The emission of direct photons from nuclear collisions has long been
considered a good probe of the temperatures and energy densities
achieved.  The reason is that the emission rate for photons of energy
$E$ is proportional to the Boltzmann factor $\exp(-E/T)$, making
photons quite sensitive to the temperature.  On the other hand, the
absolute number is obtained by integrating over space and time, so
large volumes and long lifetimes can compensate for low temperatures.
In addition, the rate quoted is in the local rest frame of the hot
matter.  If this matter is flowing, the photons can be red shifted to
lower energy or blue shifted to higher energy.  The transverse
momentum spectra of photons involves convoluting all of these effects.

\begin{figure}[t]
 \setlength{\epsfxsize}{5.5in}
  \centerline{\epsffile{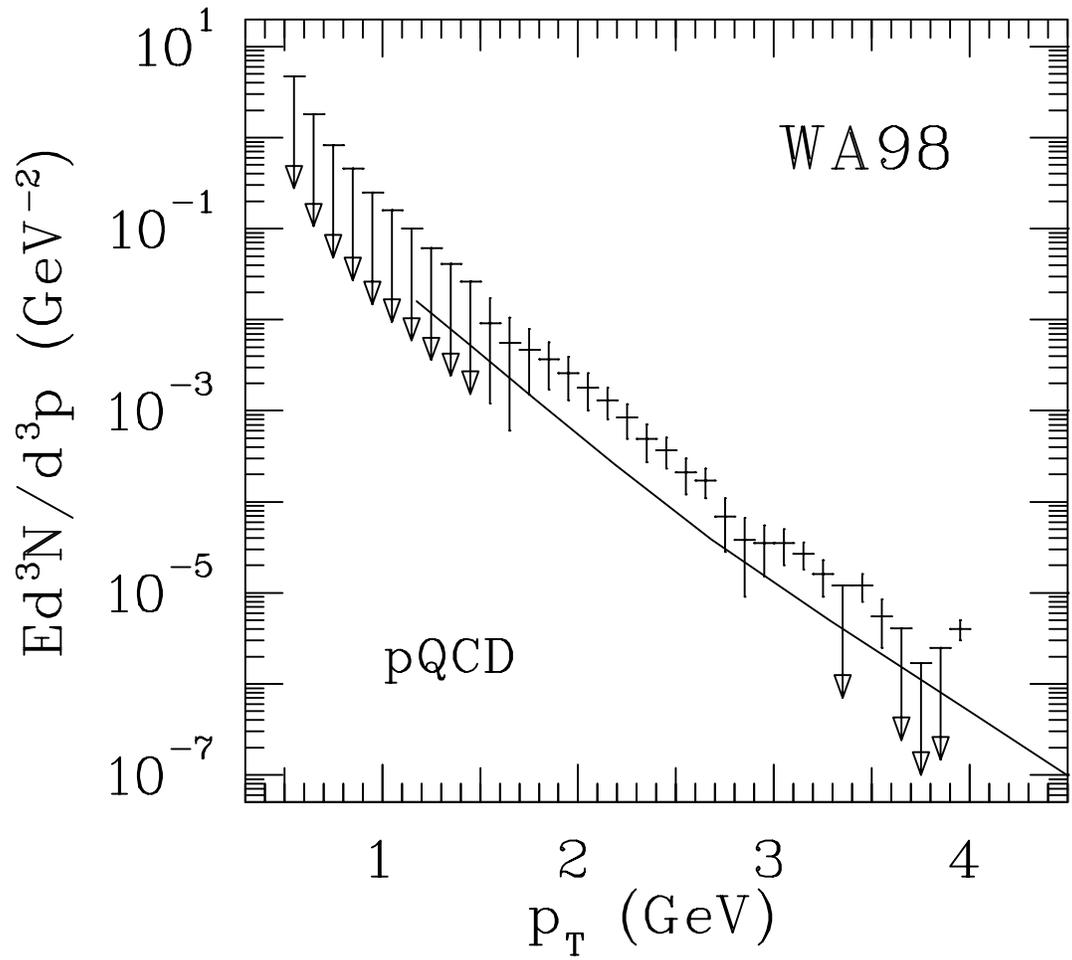}}
\vskip1cm
\caption{Photon spectrum from Pb-Pb collisions at 158 A GeV by the WA98 
         collaboration \protect\cite{wa98} compared to a perturbative
         QCD calculation \protect\cite{Wong}.}
  \label{fig9}
\end{figure}

To see whether there is a signal of thermal photon emission is very
difficult experimentally.  Model calculations have shown that thermal
emission is expected to be overwhelmed by the decays of hadrons, most
notably $\pi^0 \rightarrow\gamma\gamma$.  A lengthy and careful
analysis of Pb-Pb collisions at 158 A GeV by the WA98 collaboration
\cite{wa98} has produced the photon spectrum shown in
Fig. \ref{fig9}. All known hadronic decays into photons have been
subtracted.  This spectrum is then identified with direct photons
only. The curve in the figure is the next-to-leading order pQCD
(perturbative QCD) result of photon production from the primary
interactions of incoming partons~\cite{Wong} as shown
in~\cite{Gale01}.  In this calculation an intrinsic parton momentum of
$\langle k_T^2\rangle = 0.9$ GeV$^2$ necessary to describe
proton-induced reactions at a similar energy is assumed\footnote{It is
shown in~\cite{Dumitru} that if an additional nuclear broadening
$\Delta k_T^2 \simeq 0.5$ -- 1 GeV$^2$ is added by hand, the WA98
photon spectrum can be explained above $p_t = 2.5$ GeV.  However,
prompt photons fail to reproduce the data at low $p_t < 2.5$ GeV
regardless of the amount of broadening employed.}.  Since the photons
from primary interactions cannot explain the measured photon yield,
the yield must be associated with the secondary interactions of the
produced particles.  In the following we use the pQCD results shown in
Fig. \ref{fig9}.  It must be kept in mind that the true pQCD photons
could be less than this by a factor of two \cite{Gale01}.

\begin{figure}[t]
 \setlength{\epsfxsize}{5.5in}
  \centerline{\epsffile{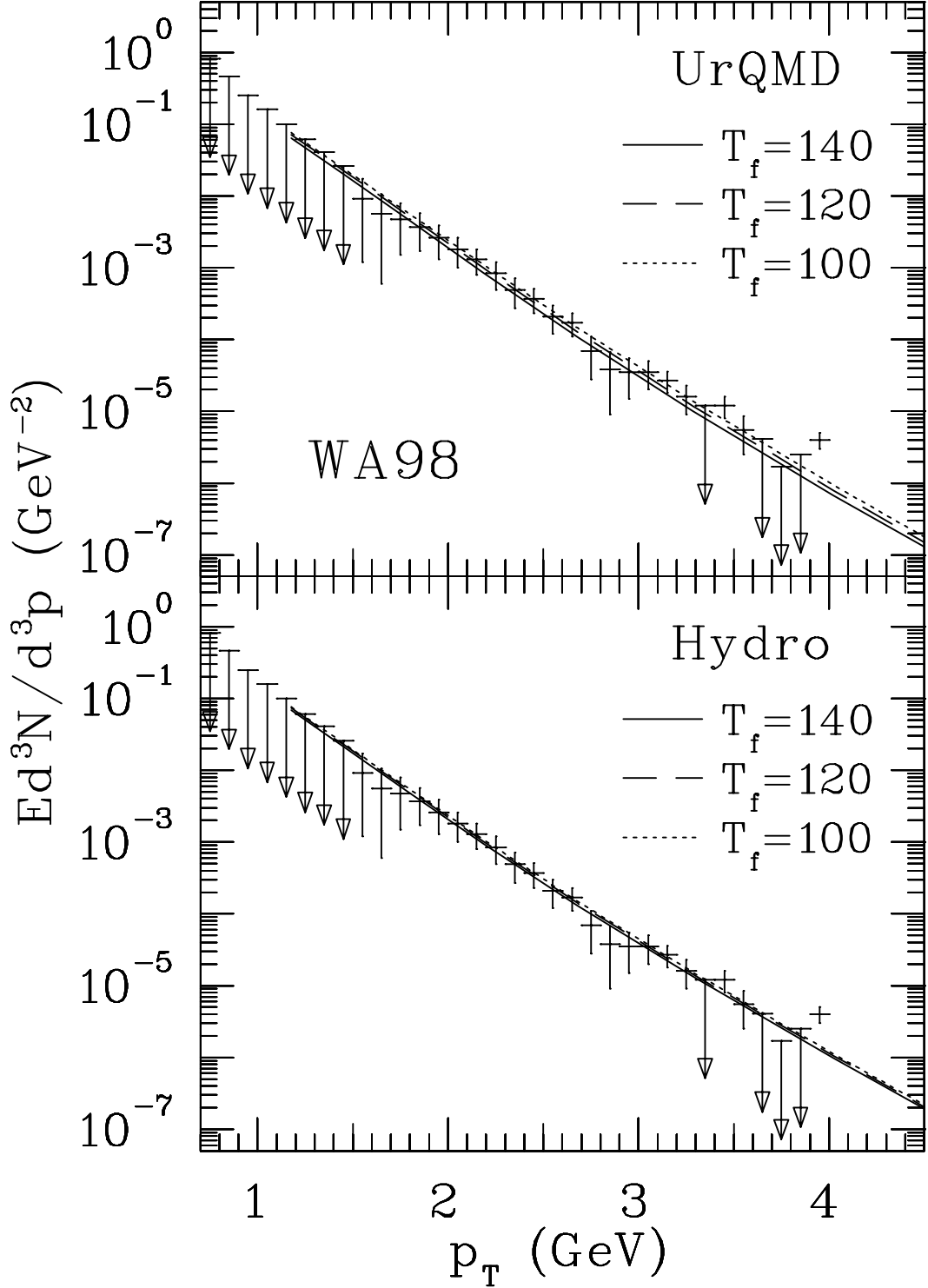}}
\vskip1cm
\caption{Comparison of the WA98 photon spectrum \protect\cite{wa98} to the
         predictions of the UrQMD model (upper panel) and the
         hydrodynamic model (lower panel) at several freeze-out
         temperatures.}
  \label{fig10}
\end{figure}

The comparison between thermal photon emission from hydrodynamic
calculations and the WA98 data has been done several
times~\cite{S2,hrr}; for a review see Ref.~\cite{pt}, for a study of
the sensitivity to the initial conditions and the equation of state
see Ref.~\cite{hrr}. In our calculations we have applied the rate
calculations of Refs.~\cite{kls,nk,xsb} in the hadronic phase and the
recent complete order-$\alpha_s$ calculation of Ref.~\cite{arn} in the
quark-gluon plasma phase. The results from the hydrodynamic
calculation are shown in the lower panel of Fig.~\ref{fig10}. Here the
pQCD result of Fig.~\ref{fig9} is added to the thermal contribution.
The agreement can only be described as excellent. Further, the results
show almost no dependence on the freeze-out temperature. As shown in
Ref.~\cite{hrr} the reason is that the contribution from the early hot
stages of the evolution dominate.

The analogous comparison using the space-time evolution of
coarse-grained UrQMD is shown in the upper part of Fig. \ref{fig10}.
Again there is hardly any dependence on the freeze-out temperature
because the emission from the early hot stage dominates.  The photon
emission has also been calculated using UrQMD on a microscopic level
\cite{dumph}. A precise comparison with our results cannot be made
since the photon production processes are not exactly the same in both
cases, nevertheless the yields are quite similar. Even so we caution
the reader regarding difficult points in our approach.  Firstly, we
began the calculation of the thermal photon emission from UrQMD at a
proper time of 1 fm/c after complete overlap. There is uncertainty
within UrQMD as to how to compute at earlier times when the strings
are still fragmenting into hadrons, or viewed differently, how to
compute photon emission before the hadrons materialize and are allowed
to interact strongly (pre-formation time).  This is very subtle, and
involves the transition from pQCD reactions to those hadronic
interactions included in UrQMD.  Secondly, as seen in Fig. \ref{fig1},
the initial temperatures can exceed 200 MeV, but the parametrization
of the thermal production rates \cite{nk} we employ were fit only in
the range $100<T<200$.  Finally, the pQCD results employed here may
overestimate the direct photons. These remarks serve to emphasize the
sensitivity of the photon yield to the early high energy density
stages of the collision.

\section{Dilepton Spectra}

\begin{figure}[t]
 \setlength{\epsfxsize}{5.5in}
  \centerline{\epsffile{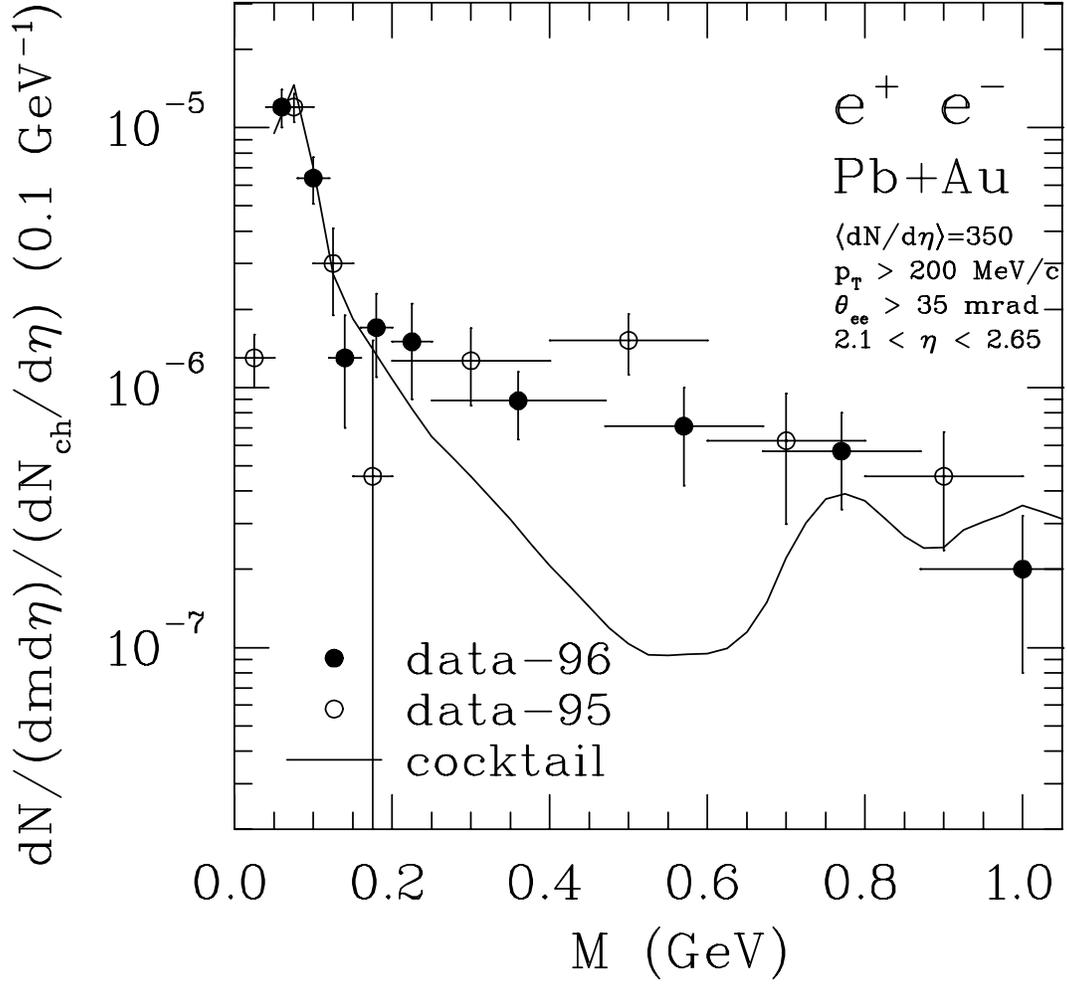}}
\vskip1cm
\caption{Comparison of the dilepton data for Pb-Au collisions at 158 A 
         GeV ('95 data Ref. \protect\cite{voi}, '96 data
         Ref. \protect\cite{len}) with the contribution from the decay
         of hadrons after freezeout.}
  \label{fig11}
\end{figure}

The dilepton spectrum measured in Pb-Au collisions is shown in
Fig. \ref{fig11}. Instead of the data shown by the CERES 
collaboration \cite{Ceres} we use the preliminary '95 and '96 data for
the most central collisions from the theses of Voigt \cite{voi} and
Lenkeit \cite{len}, respectively. These collisions have the
highest energy density which should enhance the interesting 
physics; this is indeed found to be the case experimentally \cite{len}.
It is also necessary to focus on central collisions since our
hydrodynamical model assumes cylindrical symmetry and it is
tuned to reproduce hadronic data from the most central collisions 
\cite{Jones}.  

This data has generated much excitement since it displays two very
interesting features.  The first is qualitative: there is no hint of a
peak in the $\rho$-$\omega$ mass region.  The second is quantitative:
there is a large excess of pairs in the mass range $0.25 < M < 0.7$
GeV when compared to expectations based on the decays of hadrons after
freezeout, called the ``cocktail" or ``background" contribution, and
represented by the curve in Fig. \ref{fig11}.  We use the background
estimate from Ref. \cite{len} in all of our calculations here.

\begin{figure}[t]
 \setlength{\epsfxsize}{5.5in}
  \centerline{\epsffile{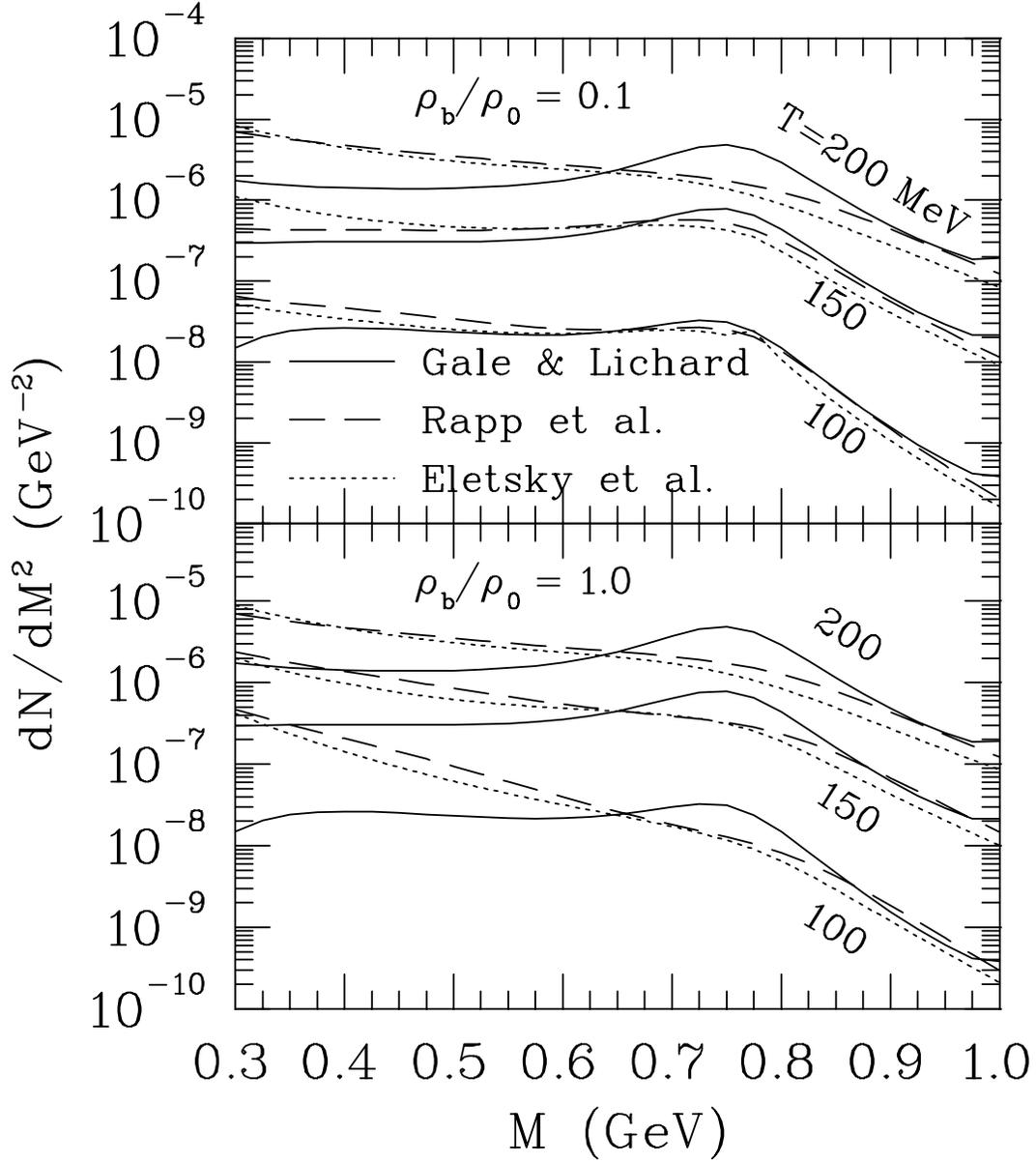}}
\vskip1cm
 \caption{Thermal dilepton emission rates computed by Gale and Lichard
         \protect\cite{galel,lich}, Rapp {\it et al.}
         \protect\cite{rapp} and Eletsky {\it et al.} 
         \protect\cite{eletsky} at various
         temperatures.  The baryon densities are fixed at 1/10 (upper
         panel) and 1 (lower panel) times the equilibrium density of 
         cold nuclear matter.}
  \label{fig12}
\end{figure}

The thermal emission rate of dileptons can be computed using many-body
theory and the vector meson dominance model for temperatures where the
matter can be described in terms of hadronic degrees of freedom.  The
most complete calculation using medium-unmodified hadrons was done by
Gale and Lichard \cite{galel,lich} and is shown by the solid line in
Fig. \ref{fig12}.  The two most sophisticated calculations using
medium-modified hadrons are shown in the same figure.  The
calculations performed by Rapp {\it et al.} \cite{rapp} are based on
solving one loop self-energy diagrams self-consistently for the
$\rho$-$\pi$-$N$ complex with interactions determined by low energy
physics.  The calculations by Eletsky {\it et al.}
\cite{eletsky} are based on scattering amplitudes for the same complex 
constructed from experimental data, namely, resonances at low energy
and Regge amplitudes at high energy.  Although the latter two
calculations are technically quite different, they aim at the same
physics, and the thermal emission rates indicated by the dashed and
dotted curves in Fig. \ref{fig12} are also very similar.  This is
reassuring; it is very important to understand these rates
quantitatively in order to draw conclusions from experiments performed
by the CERES collaboration and other groups.  Both of the
medium-modified calculations exhibit the behavior indicated by the
CERES data: suppression of the $\rho$-$\omega$ peak and enhancement at
invariant masses below 650 MeV.  The reason is that the $\rho$ and
$\omega$ mesons become greatly broadened at high energy density, with
their strength distributed mostly to lower masses. In the following we
use the rates from Eletsky {\it et al.}, although the rates of Rapp
{\it et al.} would yield essentially the same conclusions.

\begin{figure}[t]
 \setlength{\epsfxsize}{5.5in}
  \centerline{\epsffile{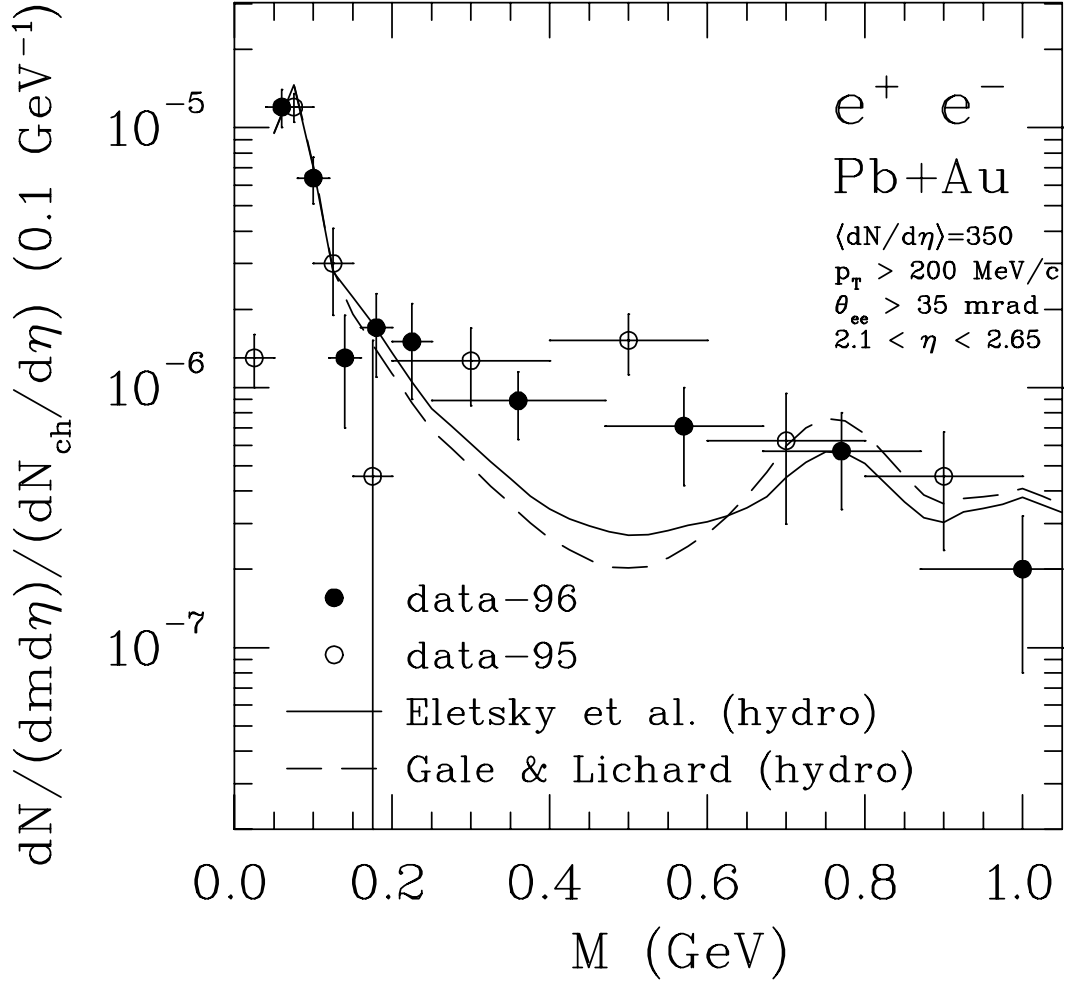}}
\vskip1cm
\caption{Comparison of the dilepton data \protect\cite{voi,len} with 
         hydrodynamic predictions using the emission rates of Gale and
         Lichard \protect\cite{galel,lich} (dashed curve) and Eletsky 
         {\it et al.} \protect\cite{eletsky} (solid curve).  The latter 
         employs medium-modified hadrons.}
 \label{fig13}
\end{figure}

As with photons, the thermal dilepton radiation from hydrodynamics has
been compared to the CERES data several times
\cite{rwrev,pasprak,rapp}.  Fig. \ref{fig13} shows our results, which
are computed for Pb-Pb collisions with a freeze-out temperature of 140
MeV. Compared to experiment the medium-unmodified thermal rate (dashed
curve) predicts somewhat fewer dileptons in the mass range from
0.3--0.6 GeV, as in Ref. \cite{pasprak}.  Some improvement is gained
by using the medium-modified rate (solid curve).

\begin{figure}[t]
 \setlength{\epsfxsize}{5.5in}
  \centerline{\epsffile{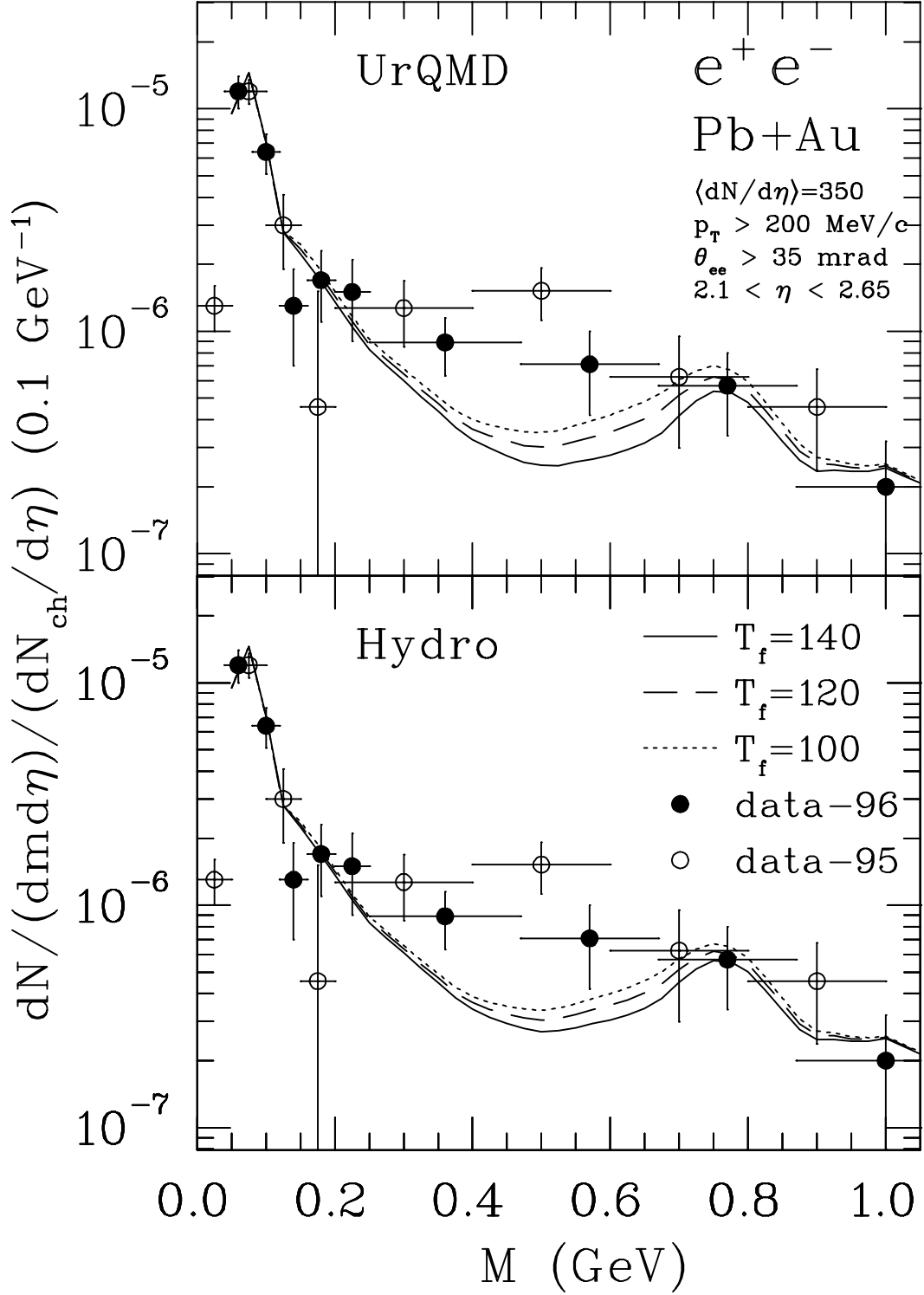}}
\vskip1cm
\caption{Comparison of the dilepton data \protect\cite{voi,len} with 
         predictions of the UrQMD model (upper panel) and the
         hydrodynamic model (lower panel) at several freeze-out
         temperatures.}
  \label{fig14}
\end{figure}

Using only the medium-modified rate, and lowering the freeze-out
temperature to 120 or 100 MeV helps fill in the intermediate mass
region a little, but not quite enough, as shown in the bottom panel of
Fig. \ref{fig14}. Going to a lower freeze-out temperature may seem to
be undesirable since it tends to enhance the $\rho$-$\omega$ mass
region and since it is inconsistent with the value determined from the
hadronic spectra.  On the other hand, there can still be contributions
to dilepton production in the free-streaming hadronic stage.  As shown
in Ref. \cite{Dinesh}, if dileptons from those few collisions in the
free-streaming stage are added to those coming from the pre-freeze-out
stage, the full result is insensitive to the numerical value of the
freeze-out temperature.  Since we do not attempt to add the
free-streaming contribution here, it is perhaps natural to use a
somewhat lower freeze-out temperature for the dileptons than for the
hadrons.

The upper panel of Fig. \ref{fig14} shows the results obtained with
the coarse-grained UrQMD. They are almost identical to the
hydrodynamical results. The slower cooling exhibited by UrQMD might
lead one to expect larger dilepton yield, but the difference in
cooling rate is too small for any significant difference to build up.
However, this effect can be seen when different freeze-out
temperatures are compared. UrQMD cools more slowly, especially during
the later stages of evolution, and therefore decreasing the freeze-out
temperature increases the lepton yield more in the UrQMD than in the
hydrodynamical model. It is also worth noticing that for the momenta
discussed here the dilepton yields, unlike the photon yields, are not
dominated by the initial hot stage of the evolution. In fact, dilepton
emission from the hot $T>200$ MeV region is a minor correction
\cite{pasprak} to the emission from cooler regions.

\begin{figure}[t]
 \setlength{\epsfxsize}{5.5in}
  \centerline{\epsffile{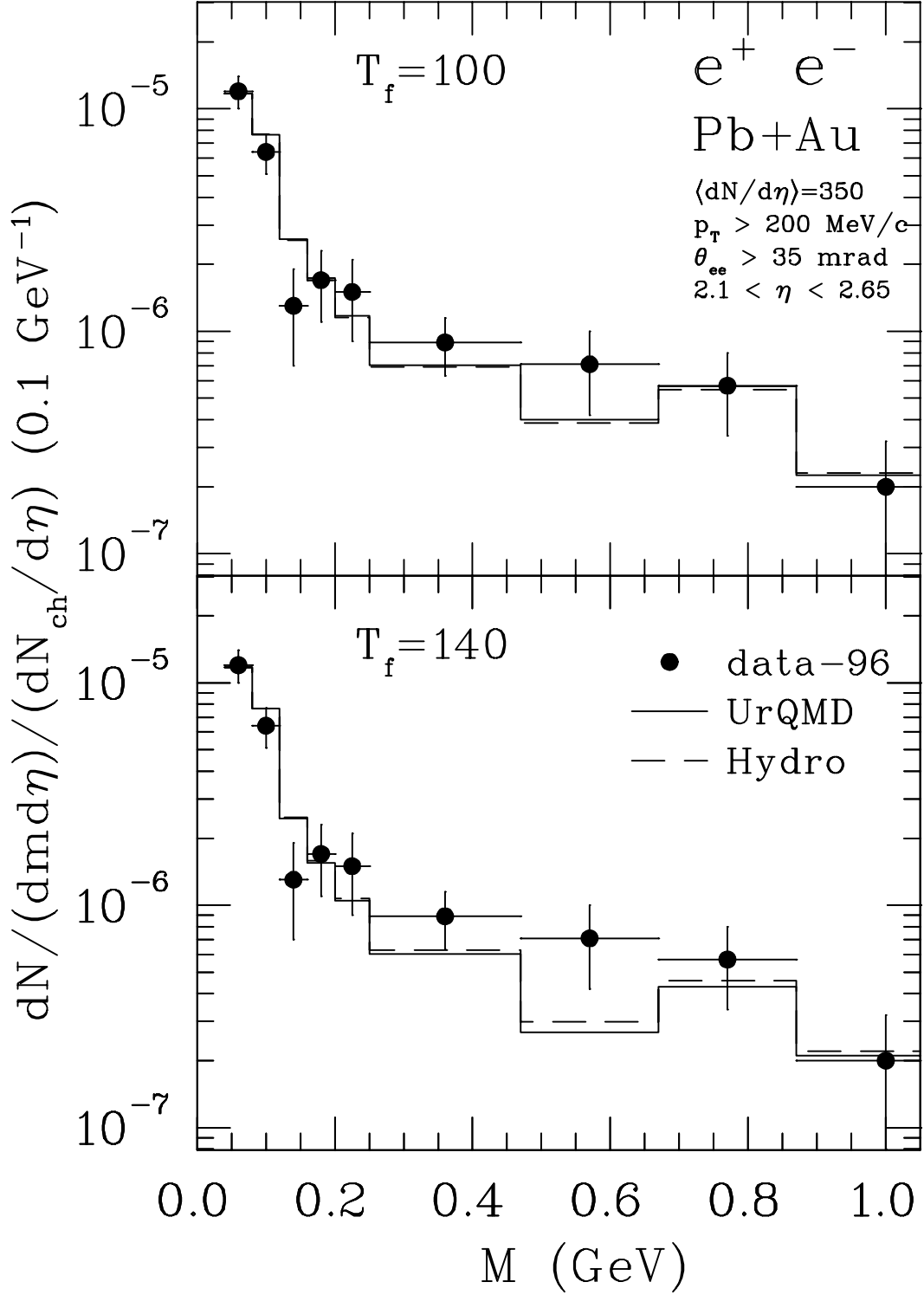}}
\vskip1cm
\caption{Comparison of the dilepton data \protect\cite{len} with 
         binned predictions of the UrQMD model (solid lines) and 
         the hydrodynamic model (dashed lines) at two freeze-out
         temperatures.}
  \label{fig15}
\end{figure}

In order to put these results in perspective we have also binned the 
calculated dilepton yields in the same way as the experimental data.
The results are shown in Fig. \ref{fig15} and compared
to just the '96 data \cite{len}. Both the UrQMD and the hydrodynamic 
results are within one standard deviation of the data when $T_f=100$ MeV. 
The agreement is only slightly worse when $T_f = 140$ MeV.
From this viewpoint it would seem premature to decide that there is a
discrepancy with theory in the 0.3--0.6 GeV mass range until
additional data are available.

There are other calculations of thermal dileptons
\cite{pasprak,rapp,rappwam}, but, with the exception of
Ref. \cite{pasprak}, these calculations were not compared to the most
central Au+Pb data.  They used data averaged over a larger range of
impact parameter, corresponding to $\langle dN/d\eta\rangle = 250$,
whereas we compare to data with $\langle dN/d\eta\rangle = 350$.  As
mentioned before, the many-body effects increase with increasing
centrality to a significant degree, particularly in the 0.3--0.6 GeV
mass range.  Further the results obtained by Rapp {\it et al.}
\cite{rapp,rappwam} were obtained with a less sophisticated approach
than employed here.  They used a very simple parametrizations of the
space-time evolution of the temperature and flow velocity, essentially
spherical fireballs or cylindrical firetubes. They did not compare
with the hadronic data, which we view as an important consistency
check.

\begin{figure}[t]
 \setlength{\epsfxsize}{5.5in}
  \centerline{\epsffile{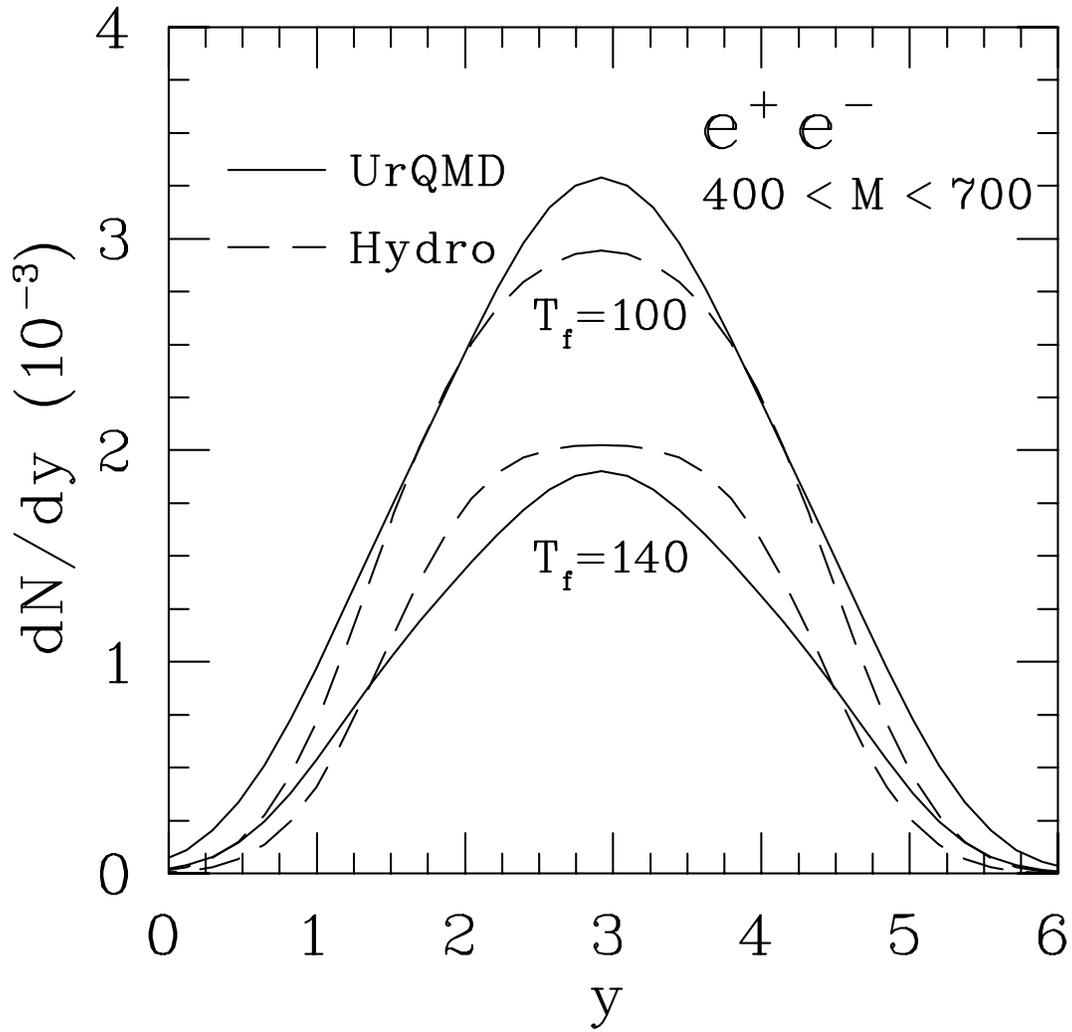}}
\vskip1cm
\caption{Rapidity distributions of dileptons in the UrQMD model (solid 
         curves) and hydrodynamic model (dashed curves) for two
         freeze-out temperatures.}
  \label{fig16}
\end{figure}

To see if the difference in initial longitudinal velocity
profile between UrQMD and hydrodynamics (Fig.~\ref{fig5}) leaves any 
observable signal we calculate the rapidity distributions of dileptons 
without any acceptance cuts, and display the results in Fig.~\ref{fig16}.
The difference in evolution leaves hardly any trace. Changing the freeze-out 
temperature has a larger effect. The small plateau in the distribution from
hydrodynamics when $T_f=140$ MeV has its origin in the first order
phase transition and the mixed phase. However, the subsequent evolution
smooths this structure away as can be seen in the distribution with
$T_f=100$ MeV. 

\begin{figure}[t]
 \setlength{\epsfxsize}{5.5in}
  \centerline{\epsffile{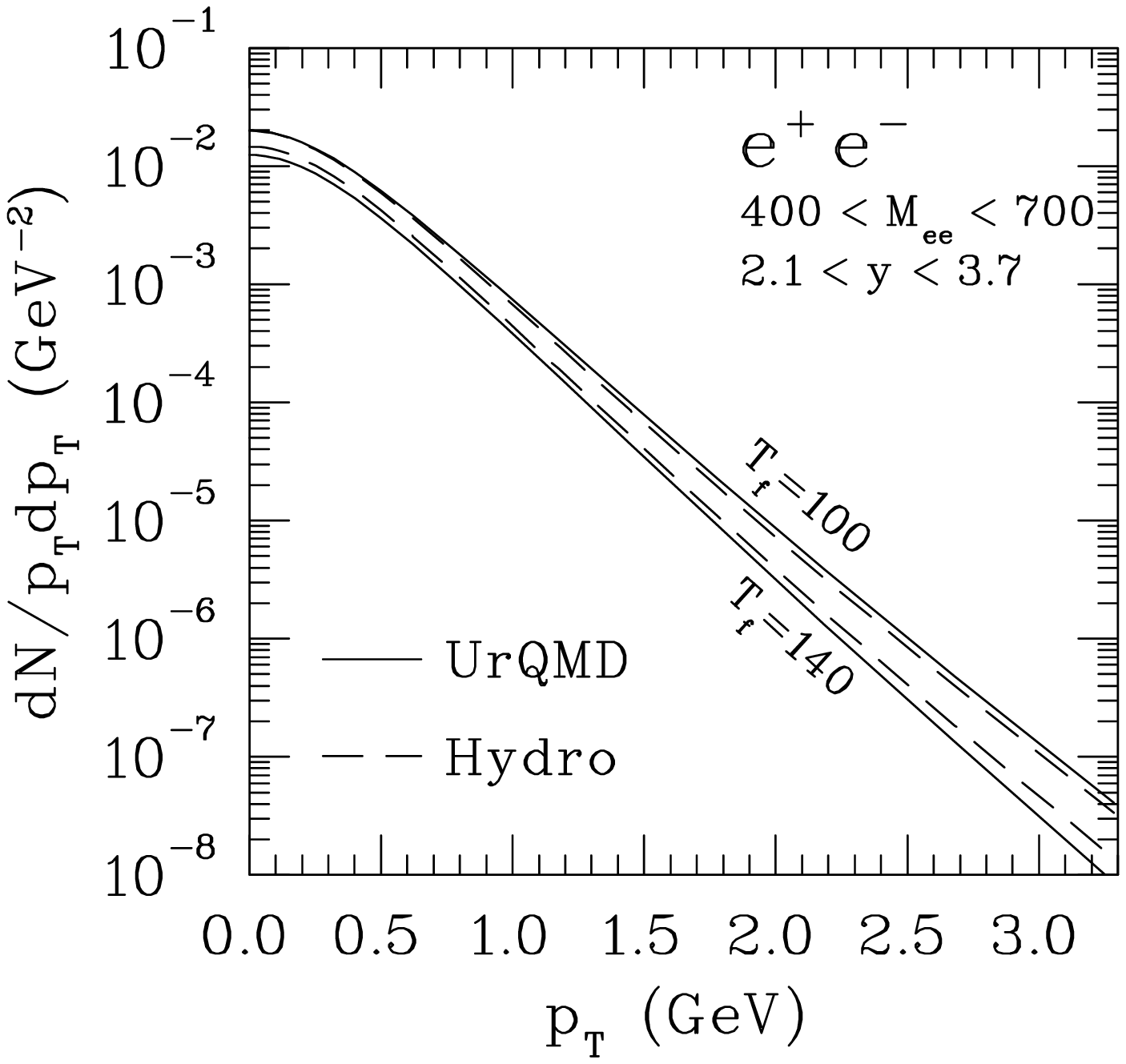}}
\vskip1cm
 \caption{The transverse momentum distributions at mid-rapidity in the 
          UrQMD model (solid curves) and the hydrodynamic model (dashed
          curves) for two freeze-out
          temperatures.}
  \label{fig17}
\end{figure}

The different times of emission of photons and dileptons are
manifested by comparison of the dilepton transverse momentum distribution
(Fig.~\ref{fig17}) and the photon distribution (Fig.~\ref{fig10}). 
As we have remarked, the photon yields are dominated by the early hot
stages of the evolution and show little dependence on freeze-out temperature.
On the other hand, the dilepton yields in Fig. \ref{fig17} clearly increase 
when the freeze-out temperature is lowered. The increase is more clearly 
seen in the transverse momentum distribution than in the mass distribution 
since the background is omitted in this figure. 

\section{Conclusion}

The expansion of matter described by hydrodynamics is rather similar
to the expansion described by (coarse-grained) UrQMD.  The transverse
flow velocities in the central plane are nearly identical, but the
temperature generally falls more slowly in UrQMD than in
hydrodynamics.  This is probably a consequence of the viscous nature
of matter that is not accounted for in perfect fluid dynamics.  The
biggest difference between the two models occurs for the longitudinal
flow velocity; its gradient along the beam axis is larger in UrQMD
than in hydrodynamics. Since the initial conditions, so to speak, in
UrQMD are a direct consequence of string dynamics there is no freedom
to adjust them. It would be very interesting to modify the initial
conditions in hydrodynamics to match those of UrQMD. It would also be
very interesting to include viscosity and heat conduction in the fluid
dynamical calculations to determine exactly how close the
coarse-grained UrQMD results are to viscous flow.

Taking the coarse-grained UrQMD flow, temperature, and chemical
potential fields as given, we computed the hadronic spectra in exactly
the same way as in hydrodynamics. A freeze-out temperature of about
120$\pm$10 MeV reproduces the experimental hadronic data of NA49 
reasonably well. This should not be considered a substitute for the 
original hadronic spectra from UrQMD, but only serves to check that 
the coarse-graining procedure has not severely distorted the results.
Both hydrodynamics and UrQMD suggest that the matter created has a
local temperature between 120 and 250 MeV for a period of about 15
fm/c with a transverse extent of about 10 fm.

The direct photon spectrum of WA98 is very well reproduced by a
combination of perturbative QCD for the first hard scattering plus
thermal radiation from either a hydrodynamic or an UrQMD expansion.
It must be kept in mind that the pQCD calculation of direct photons 
is uncertain. In any case the high transverse momentum photons are 
sensitive to the early high-energy-density stage of the collision.

Both hydrodynamics and coarse-grained UrQMD give very similar numbers
of low mass dileptons. The thermal emission rates seem to be rather
robust since the different computations by Rapp {et al.} \cite{rapp}
and Eletsky {\it et al.} \cite{eletsky} give such similar results,
although some uncertainty remains in the background ``cocktail"
contribution.  These calculations tend to fall slightly below the data
for the most central collisions in the invariant mass range 0.3--0.6
GeV, but the effect is less evident if the results are binned to
conform to the experimental analysis. We suggest that further data are
needed in order to decide whether a significant discrepancy with
theory exists.

There is no doubt that the direct photon and low mass dilepton
measurements require the formation of high density matter.  Further
data from central collisions at the SPS would be very useful and
welcome.  We also await the first results from RHIC on electromagnetic
probes.

\section*{Acknowledgments}

This work was supported by the US Department of Energy grant 
DE-FG02-87ER40328.

\setlength{\topmargin}{-1cm}
\setlength{\textheight}{25cm}

\newpage

\newpage

\newpage

\newpage

\newpage

\newpage

\newpage

\newpage

\newpage

\newpage

\newpage

\newpage

\newpage

\newpage

\newpage

\newpage

\begin{thebibliography} {40}

\bibitem{urqmd1} 
 S. A. Bass, M. Belkacem, M. Bleicher, M. Brandstetter, L. Bravina, C. Ernst, 
 L. Gerland,  M. Hofmann, S. Hofmann, J. Konopka, G. Mao, L. Neise, S. Soff,
 C. Spieles, H. Weber, L. A. Winckelmann, H. St\"ocker, W. Greiner, 
 Ch. Hartnack, J. Aichelin and N. Amelin,
 Prog. Part. Nucl. Phys. {\bf 41}, 225 (1998).

\bibitem{urqmd2}
 M. Bleicher, E. Zabrodin, C. Spieles, S. A. Bass, C. Ernst, S. Soff, L. 
 Bravina, M. Belkacem, H. Weber, H. St\"ocker and  W. Greiner, J. Phys. G: 
 Nucl. Part. Phys. {\bf 25}, 1859 (1999).

\bibitem{soll} J. Sollfrank, P. Huovinen, M. Kataja, P. V. Ruuskanen, M. 
Prakash and R. Venugopalan, Phys. Rev. C {\bf55}, 392 (1997).

\bibitem{Sollfrank99} J. Sollfrank, P.~Huovinen and P. V. Ruuskanen,
 Eur.\ Phys.\ J.\ C {\bf 6}, 525 (1999).

\bibitem{pt} T. Peitzmann and M. H. Thoma, hep-ph/0111114.

\bibitem{rwrev} R. Rapp and J. Wambach, Adv. Nucl. Phys. {\bf25}, 1 (2000).

\bibitem{mohamed} M. Belkacem, M. Brandstetter, S. A. Bass, M. Bleicher,
 L. Bravina, M. I. Gorenstein, J. Konopka, L. Neise, C. Spieles, S. Soff,
 H. Weber, H. St\"ocker and W. Greiner, Phys. Rev. C {\bf58}, 1727 (1998).

\bibitem{Huovinen99} P. Huovinen, P. V. Ruuskanen and J. Sollfrank,
 Nucl.\ Phys.\ A {\bf 650}, 227 (1999).

\bibitem{pasprak} P. Huovinen and M. Prakash, 
 Phys. Lett. B {\bf450}, 15 (1999).

\bibitem{urhad} M. Bleicher, C. Spieles, C. Ernst, L. Gerland, S. Soff,
 L. Neise, H. St\"ocker, W. Greiner and S. A. Bass, 
 Phys. Lett. B {\bf447}, 227 (1999).

\bibitem{urstr} S. Soff, D. Zschiesche, M. Bleicher, C. Hartnack, 
 M. Belkacem, L. Bravina, E. Zabrodin, S. A. Bass, H. St\"ocker
 and W. Greiner, J. Phys. G: Nucl. Part. Phys. {\bf27}, 449 (2001).

\bibitem{Jones} P. G. Jones {\it et al.}  [NA49 Collaboration],
  Nucl.\ Phys.\ A {\bf 610}, 188c (1996).

\bibitem{wa98} M. M. Aggarwal {\it et al.} [WA98 Collaboration], Phys. Rev. 
 Lett. {\bf85}, 3595 (2000).

\bibitem{Wong} C. Y. Wong and H. Wang,
 Phys.\ Rev.\ C {\bf 58}, 376 (1998).

\bibitem{Gale01} C. Gale,
 Nucl.\ Phys.\ A {\bf 698}, 143c (2002).

\bibitem{Dumitru}
 A.~Dumitru, L. Frankfurt, L. Gerland, H. St\"ocker and M. Strikman,
 Phys. Rev. C {\bf 64}, 054909 (2001).

\bibitem{S2} D. K. Srivastava and B. Sinha,
 Phys. Rev. C {\bf 64}, 034902 (2001);
 J. Alam, S. Sarkar, T. Hatsuda, T. K. Nayak and B. Sinha,
 {\it ibid.} {\bf 63}, 021901 (2001);
 J. Alam, P. Roy, S. Sarkar and B. Sinha, nucl-th/0106038;
 D. Y. Peressounko and Y. E. Pokrovsky, hep-ph/0009025;
 A. K. Chaudhuri, nucl-th/0012058;
 K. Gallmeister, B. Kampfer and O. P. Pavlenko,
 Phys. Rev. C {\bf 62}, 057901 (2000);
 F. D. Steffen and M. H. Thoma, Phys. Lett. B {\bf 510}, 98 (2001).

\bibitem{hrr} P. Huovinen, P. V. Ruuskanen and S. S. R\"as\"anen,
 nucl-th/0111052.

\bibitem{kls} J. Kapusta, P. Lichard and D. Seibert, Phys. Rev. D {\bf 44}, 
 2774 (1991).

\bibitem{nk} H. Nadeau, J. Kapusta and P. Lichard, Phys. Rev. C {\bf 45}, 
 3034 (1992). 

\bibitem{xsb} L. Xiong, E. Shuryak and G. E. Brown, Phys. Rev. D {\bf46}, 
 3798 (1992).

\bibitem{arn} P. Arnold, G. D. Moore and L. G. Yaffe, 
 JHEP {\bf0111}, 057 (2001); JHEP {\bf 0112} (2001).

\bibitem{dumph} A. Dumitru, M. Bleicher, S. A. Bass, C. Spieles, L. Neise,
 H. St\"ocker and W. Greiner, Phys. Rev. C {\bf57}, 3271 (1998).

\bibitem{Ceres} B.~Lenkeit {\it et al.}  [CERES Collaboration],
 Nucl. Phys. A {\bf 661}, 23 (1999).

\bibitem{voi} C. Voigt, Doctoral Thesis, University of Heidelberg, 1998.

\bibitem{len} B. Lenkeit, Doctoral Thesis, University of Heidelberg, 1998.

\bibitem{galel} C. Gale and P. Lichard, Phys. Rev. D {\bf49}, 3338 (1994).

\bibitem{lich} P. Lichard, Phys. Rev. D {\bf 49}, 5812 (1994).

\bibitem{rapp} R. Rapp, G. Chanfray and J. Wambach,
 Nucl. Phys. {\bf A617}, 472 (1997);
 R. Rapp, Phys. Rev. C {\bf 63}, 054907 (2001).

\bibitem{eletsky} V. L. Eletsky, M. Belkacem, P. J. Ellis and J. I. Kapusta, 
 Phys. Rev. C {\bf 64}, 035202 (2001).

\bibitem{Dinesh}
 D. K. Srivastava and J. I. Kapusta, Phys. Rev. C {\bf 48}, 385 (1993).

\bibitem{rappwam} R. Rapp and J. Wambach, Eur. Phys. J A {\bf 6}, 415 (1999).

\end{thebibliography}
\end{document}